%% file: wwv.tex
\documentclass{eptcs}
\providecommand{\event}{WWV 2012\\Automated Specification and Verification of Web Systems 8th International Workshop (as part of DisCoTec'12)}
\providecommand{\titlerunning}{The Jasper Framework}
\providecommand{\authorrunning}{James Smith}
\usepackage{breakurl}             

\usepackage{myarray}
\usepackage{myspace}
\usepackage{myletters}
\usepackage{myabstractlisting}

\usepackage{url}
\usepackage{float}
\usepackage{amsmath}
\usepackage{amssymb}
\usepackage{caption}
\usepackage{listings}
\usepackage{graphicx}

\title{The Jasper Framework: Towards a Platform Independent, Formal Treatment of Web Programming}
\author{James Smith
\institute{Imperial College\\ London, United Kingdom}
\email{jecs@imperial.ac.uk}
}

\begin{document}
\maketitle

\setlength\abovecaptionskip{0pt}
\setlength\belowcaptionskip{0pt}

\captionsetup[figure]{name=Listing}
\numberwithin{figure}{section}

\begin{abstract}
This paper introduces Jasper, a web programming framework which allows web applications to be developed in an essentially platform indepedent manner and which is also suited to a formal treatment. It outlines Jasper conceptually and shows how Jasper is implemented on several commonplace platforms. It also introduces the Jasper Music Store, a web application powered by Jasper and implemented on each of these platforms. And it briefly describes a formal treatment and outlines the tools and languages planned that will allow this treatment to be automated.
\end{abstract}

\input{introduction}
\input{preliminaries}
\input{core}

\input{extending}

\input{forms}

\input{implementation}
\input{jms}

\input{formal}

\input{related}
\input{conclusions}

\nocite{*}
\bibliographystyle{eptcs}
\bibliography{intro}

\end{document}

%% file: introduction.tex
\section{Introduction}


These days a web application may span both server and client side. Facebook is a good example, with much of the site's functionality being implemented using JavaScript and executed in the browser. Communication between client and server sides is also, typically, not just characterised by requests from the browser, requests are also made by JavaScript in the form of Ajax calls. It is not at all obvious, for example, that Facebook fakes conventional browser requests. Clicking on menu items appears to result in such conventional requests being made, since the whole page seems to change, but in fact these are often Ajax calls and the whole page is changed dynamically, rather than a new page being served. In this way the chat pane can remain ``on top'', while the page changes ``underneath''. Such techniques used to characterise what were known as rich web applications but they are becoming so commonplace nowadays that a rich web application might be considered the norm.

The Jasper framework does not encompass such rich web applications, at least not directly. It is solely a server side framework. This is not to say that a Jasper web application cannot support Ajax calls. These calls are no more than HTTP requests, after all. However, considering web application programming from a Jasper point of view does not, at least not directly, include consideration of code executed on the client side. Handling HTTP requests regardless of where they originate and generating responses is primarily what Jasper is about.

Looking at the most commonplace languages used in server side programming today it is clear that there are as many similarities as their are differences between them. Syntactic differences hide the fact that for non-specific purposes their operation is practically the same. Furthermore, if the problem to be solved does not change across platforms, there is good reason to expect that the respective solutions bear similarities, too. Jasper's approach therefore is to distill patterns inherent in the operation of simple web applications into patterns and techniques that are not tied to any particular platform.

The contributions of this paper therefore are: firstly, to introduce the Jasper web programming framework; secondly, to demonstrate by way of concrete implementations that it is possible to approach web programming in an essentially platform indepedent manner; and thirdly, to show that it is possible to reason formally about these types of web applications.

\subsection{Resources}
Jasper has a dedicated website~\cite{jasper} which includes an API reference, tutorials, and more. There is also a demonstration site, called the Jasper Music Store~\cite{jms}. The source distributions for both the Jasper API and Jasper Music Store can be downloaded from the Jasper web site. In the interests of brevity many listings have been omitted from this paper. The online API documentation~\cite{jasper-api} contains listings of the core API classes, however.

\subsection{How this paper is organised}
After this introdution, the preliminaries section defines the notions of server context and active content, then gives a brief overview of Jasper's components. The next three sections explain Jasper in detail. The implementation section details running a Jasper website on each of the supported platforms and touches on configuration issues. The section on the Jasper Music Store explains how this demonstration site attains a suprising degree of platform independence. The formal treatment section outlines very briefly the theory related to Jasper. There then follows a related work section. Finally, the conclusions and future work section is followed by the acknowledgements.

%% file: preliminaries.tex
\section{Preliminaries}
\subsection{Web application contexts and active content}
Typically the interaction between a web browser and a web server is carried out by way of the HTTP protocol. Consider the following URL:
\begin{lstlisting}[basicstyle=\ttfamily\footnotesize]
http://www.examplesite.com/test/index.html
\end{lstlisting}
Typing this URL into a browser address bar instructs the browser to send an HTTP GET request to the www.examplesite.com website. Such a slimmed down request looks like this:
\begin{lstlisting}[basicstyle=\ttfamily\footnotesize]
GET /test/index.html HTTP/1.1
\end{lstlisting}
Literally, the browser asks ``get me the resource located at /test/index.html''. In order to serve the resource, the server must find it on its file system. To do this, the resource URI is broken down into the web application path and the resource name, in this case `/test/' and `index.html', respectively. The web application path is then mapped to a location on the file system, `/home/usr/www/test', for example. This mapping, or ordered pair \texttt{(web application path, server file system location)} is called the server context. This is always set, in a variety of ways, for any web application, on any platform. 

The content served by the mechanism outlined above is known as static content. In these cases resources are simply files, and these, once located, are delivered to the browser as-is. Most content is served actively, however, in the sense that the server software or software that it employs to deliver the content typically does one of two things:
\begin{enumerate}
\item Restricts the content, serving it only to authenticated users
\item Modifies the content in some way before returning it
\end{enumerate}
A program or script that the server invokes in order to serve content actively is called an active server process, or server process for short. A server process might be, for example, a Java servlet; a C\# DLL invoked from an ASPX page; a Perl script invoked via the CGI interface; or a PHP script. Active content should not be confused with dynamic content, which usually implies content that is generated by JavaScript executing on the client side. 

\subsection{Jasper's components}

The Jasper distribution comprises an API, example code and online documentation. The API and example code are implemented on four commonplace web programming platforms, namely Java/servlets, C\#/ASP.net, PHP and Perl/CGI.

The API consists of core and extension classes. The seven core classes include five template classes plus the Form and Config classes. Essentially these classes implement a very simple template engine. The extension classes, around a dozen in total, abstract the platform specifics to leave the core and example code identical across platforms apart from syntactic differences. There is an extension class to handle regular expressions, for example, and one to provide a basic interface to the MySQL database.

The example code comprises the source code and supporting files for a very simple web application that presents the user with a feedback form and sends an email upon this form's successful validation. The online documentation details the steps required to build this application, therefore giving a practical introduction to programming with Jasper. It also contains some platform specific notes and other helpful information.

%% file: core.tex
\section{Jasper's core classes}
This section outlines the methods and patterns in the core API and justifies them in a series of steps.

\textit{\textbf{Step 1:}Active server processes are instructed to serve a resource in much the same way that a server is instructed when serving a static resource. It is likely that the resource needs to be restricted or modified in some way, but this is beside the point. The point is, at the risk of repetition, that a server process is always required to return a specific, named resource.}

This is not always the case, but should be. PHP and Perl can be embedded within the resource itself, for example. Similarly with C\# and ASPX pages. And Java can be embedded within JSP pages. For all but the most trivial of web applications, however, this is a questionable approach, for two reasons. Firstly, embedding one language in another in the same file makes it difficult to read and maintain. In the case of ASPX pages, for example, this approach can involve a whole new set of programming objects with their associated markup. Secondly, and more fundamentally, embedding code in a resource breaks the two rules for active content given in section 2.3. You cannot easily restrict a resource if the code for doing so is embedded within it, for example. Consider a page that only gets shown if a user is authenticated. The user requests the page, which must execute, at least partially. Code within it, once appraised of the fact that the user is not authenticated, must either send an HTTP redirect to instruct the browser to load an alternate page or must suppress the content of the page somehow, perhaps by wiping the output buffer, and then instead show the content of the alternate page by some means. Both solutions are unsatisfactory. 
 
Moving on, from now on this paper will be concerned with HTML files as resources. In fact the presence of HTML markup makes no difference to the reasoning that follows, as Jasper server processes take no notice of it. To a Jasper server process, an HTML file is simply a text file. Text files are used here because they admit both restriction and modification. Binary files typically admit only the former.

\textit{\textbf{Step 2:}Active server processes modify a text file by searching for specific markup and replacing this markup with actively generated content.}

The markup can be chosen so as not to conflict with HTML or any other commonplace markup. Jasper markup is characterised by what are called tokens, which are content enclosed in a double set of square braces. Here is some example HTML with such a token:
\begin{lstlisting}[basicstyle=\ttfamily\footnotesize]
<p>The current date is <strong>[[vCurrentDate]]</strong></p>
\end{lstlisting}
This, once modified, might appear as the following:
\begin{lstlisting}[basicstyle=\ttfamily\footnotesize]
<p>The current date is <strong>10th May 2011</strong></p>
\end{lstlisting}
Note that the HTML markup remains intact. HTML or other text files with such markup are known as template files.

\textit{\textbf{Step 3:}In order to process a template file, an active server process must read it line by line, and process each line before concatenating the results and returning them.}

This step gives rise to the \texttt{process\_file()} method of the \texttt{TemplatePlain} class, which passes each line to the \texttt{process\_line()} method of the \texttt{Template} class before concatening the results and returning them. It makes use of the \texttt{File} and \texttt{FileUtils} classes, two of the Jasper API's extension classes, of which there are around ten. As mentioned in the introduction, the purpose of the extension classes is to abstract the platform specifics so that the core classes and the remainder of the classes of the web application can be written in a platform independent way.

\textit{\textbf{Step 4:}In order to process a line of a template file, an active server process must find all of the tokens in the line and process each, before concatenating the results in order, together with the content in between. The resulting string is then returned.}

This step gives rise to the aforementioned \texttt{process\_line()} method of the \texttt{Template} class. A regular expression matches tokens, the contents of which are passed to a \texttt{process\_token()} method. Native functionality is again wrapped in extension classes, the \texttt{RegEx} and \texttt{RegExResult} classes in this case.

Before taking any more steps, list files are introduced. Since web applications often have to form query strings, list files provide a way to make them readable, and to form them actively. Consider the following list file:
\begin{lstlisting}[basicstyle=\ttfamily\footnotesize]
name=[[name]]
message=[[message]]
\end{lstlisting}
If the tokens are replaced with the appropriate active content, the content either side of the equals sign escaped, and then the lines concatenated and separated by ampersands, the following query string might result:
\begin{lstlisting}[basicstyle=\ttfamily\footnotesize]
name=James+Smith&message=Hello,+world!!
\end{lstlisting}

This functionality is implemented by the \texttt{process\_file()} method of the \texttt{TemplateList} class. In a similar vein to the \texttt{process\_file()} method of the \texttt{TemplatePlain} class, each line is passed to the \texttt{process\_line()} method of the \texttt{Template} class before being processed further.

\textit{\textbf{Step 5:}Web applications rely on name-value pairs of textual content for configuration. Configuration files are parsed by an active server process whenever it executes. In order to make these configuration variables available to the server process, it is natural to place them in an associative array.}

From this step comes a simple configuration file format:
\begin{lstlisting}[basicstyle=\ttfamily\footnotesize]
#xSuppressSQLLogging=TRUE

gFeedbackSenderName=Site feedback
gFeedbackSubject=Site feedback

smtpHost=localhost
\end{lstlisting}

\begin{figure}[H]
\begin{lstlisting}[basicstyle={\sffamily\footnotesize},frame=single,framesep=3pt,xleftmargin=3pt,xrightmargin=3pt]
Date date = new Date();
TemplateHTML templatehtml = new TemplateHTML();

props{ "VAR.vCurrentDate" } = date.currentDate.toString();
string ret = templatehtml.processfile( props{ "CONFIG.rootDir" } +
                                           "template/_inc/" +
                                               "date" +
                                                   ".html", props );
unset( props{ "VAR.vCurrentDate" } );
\end{lstlisting}
\caption{Echoing the current date}
\label{current_date}
\end{figure}

Configuration files are processed by the \texttt{parse()} method of the \texttt{Config} class, which prepends the `CONFIG.' string to the names of the configuration variables before placing them in an associative array.

\textit{\textbf{Step 6:}Web applications should have access to form variables, unless they are uploaded files or other such binary data, by the same means that gives them access to configuration variables.}

The \texttt{parse()} method of the \texttt{Form} class is responsible for retrieving form variables and placing them in an associative array. Since these variables are usually held in the same associative array as configuration variables, this method prepends the `FORM.' string to the names of form variables, in order to differentiate between the two. The body of this method is necessarily platform specific. 

\textit{\textbf{Step 7:}Web applications should be able to add temporary variables to the same associative array that holds form and configuration variables, in order to echo them in template and list files.}

Consider again part of a template file:
\begin{lstlisting}[basicstyle=\ttfamily\footnotesize]
<p>The current date is <strong>[[vCurrentDate]]</strong></p>
\end{lstlisting}
Listing~\ref{current_date} demonstrates how a Jasper server process can process this file. The `VAR.' string is prepended to the names of temporary variables, in order to differentiate them from form and configuration variables. Temporary variables are also typically removed from the associative array immediately once they are no longer needed, hence their name. The language here is known as concrete pseudo-code, by the way, called as such because it very closely resembles the concreate languages in which Jasper is written. 

The associative array that holds configuration, form and temporary variables is passed as the last parameter to the majority of the methods in a Jasper server process and for this reason is known as the global properties array. It plays a fundamental role in reasoning formally about Jasper web applications. The \texttt{process\_token()} methods of the \texttt{TemplateHTML} and \texttt{TemplateGET} classes, identical in both cases, echo the values of temporary, form and configuration varables in that order.

%
%

To summarise, the core of Jasper's API comprises the \texttt{Form} and \texttt{Config} classes, together with the five template classes. Because the extension classes do nothing but abstract native functionality, by default Jasper does nothing more than echo the values of temporary, form and configuration variables in template and list files.

%% file: extending.tex
\section{Extending the core template hierarchy}
This section shows how the core template hierarchy can be extended to add support for custom tokens both to create hyperlinks and for form validation.

In practice there is never any need to extend Jasper's default functionality in relation to list files, in all cases simply being able to echo the values of configuration and temporary variables is enough. In the case of template files, however, it is natural to override the \texttt{process\_token()} method of the \texttt{TemplateHTML} class in order to support custom tokens.

The first custom tokens to be considered are a `[[MAIN]]' token, which results in the URI of a server process; and a `[[PAGE:...]]' token, which results in a query string to instruct a server process which template file to process. The \texttt{process\_token()} method of the \texttt{TemplateBase} class, which usually extends the \texttt{TemplateHTML} class,  implements this functionality. If the token passed to this method is not matched, the \texttt{process\_token()} method of the subclass is called. In this way, Jasper's default token functionality results. This is typical, with \texttt{process\_token()} methods nearly always calling the \texttt{process\_token()} method of their subclass should they fail to match the token themselves.

The `main.html' template file has three tokens to echo the value of configuration variables and thereby create a URI of a server process:
\begin{lstlisting}[basicstyle=\ttfamily\footnotesize]
[[appPath]][[procPath]]main[[procExt]]
\end{lstlisting}
The server process here is called `main'. A Jasper web application typically has from one to a dozen uniquely named server processes, rather than having one entry point. 

The list file to create the query string contains just one token to echo the value of the temporary variable:
\begin{lstlisting}[basicstyle=\ttfamily\footnotesize]
page=[[vPage]]
\end{lstlisting}
With the \texttt{TemplateBase} class it is now possible to add custom tokens in the following way:
\begin{lstlisting}[basicstyle=\ttfamily\footnotesize]
<a href="[[MAIN]]?[[PAGE:test]]">Test page</a>
\end{lstlisting}
Next, a token is added to include a simple feedback form. The \texttt{process\_token()} method of the \texttt{TemplateMain} class, shown in listing~\ref{TemplateMain:process_token}, which usually extends the \texttt{TemplateBase} class, processes this token. At this level, each custom token typically has an associated class. 

The template HTML for the feedback form includes several tokens:
%
%
%
%
%
%
\begin{lstlisting}[basicstyle=\ttfamily\footnotesize]
<form action="[[MAIN]]" method="POST">

    <span>Your full name:</span><span class="exclaim">[[EXCLAIM:fullname]]</span>
    <input type="text" name="fullname" value="[[fullname]]"/>

    <span>Your comments:</span><span class="exclaim">[[EXCLAIM:comments]]</span>
    <textarea name="comments">[[comments]]</textarea>

    <input type="hidden" name="page" value="[[page]]"/>
    <input type="hidden" name="command" value="FEEDBACK"/>
    <input type="submit" value="SUBMIT"/>

    <span class="error">[[ERROR]]</span>

<form/>
\end{lstlisting}

The `[[fullname]]' ,`[[comments]]', and `[[page]]' tokens echo the value of form variables. The `[[EXCLAIM:...]]' and `[[ERROR]]' tokens, on the other hand, provide functionality specific to form validation, and are implemented in the \texttt{TemplateFormErrors} class. This extends the \texttt{TemplateForm} class, which itself processes custom tokens to support radio buttons, check boxes and drop down lists. The `[[EXCLAIM:...]]' token prints an exclamation mark corresponding to an invalid form field when necessary, whilst the `[[ERROR]]' token prints the related error. Temporary variables are used to store the name of the invalid field and the associated error message.

\begin{figure}[H]
\begin{lstlisting}[basicstyle={\sffamily\footnotesize},frame=single,framesep=3pt,xleftmargin=3pt,xrightmargin=3pt]
class TemplateMain extends TemplateBase {    

    function process_token ( token, props ) {
        if ( StringUtils.equals( token, "FEEDBACK_FORM" ) ) {
            return ( FeedbackForm.process( props ) );
        }
        return ( parent.process_token( token, props ) );
    }
}
\end{lstlisting}
\caption{The \texttt{process\_token()} method of the \texttt{TemplateMain} class}
\label{TemplateMain:process_token}
\end{figure}

The \texttt{TemplateFormErrors} class is rarely if ever extended. On the other hand the \texttt{TemplateBase} class might be expanded in practice to support several server processes and the \texttt{TemplateMain} class could well, in a production web application, support over a hundred custom tokens, each corresponding to a form or other such element such as a list or a menu.

%% file: forms.tex
\section{Form handling}
This section describes preprocessing and a standard way to handle forms. It concludes with an overview of the architecture of a simple Jasper web application.

Form submissions need to be handled before a page's active content is generated because they may have a bearing on which page is shown. If a feedback form's fields are valid, for example, a web application may show a confirmation page rather than show the form again. The functionality associated with handling forms and generating active content is known as preprocessing.

Preprocessing is typically carried out by a dedicated class and its dependent classes. A suitable \texttt{PreProcess} class for handling the feedback form is shown in listing~\ref{PreProcess}. Depending on the value returned by the \texttt{preprocess()} method of the \texttt{FeedbackForm} class, the \texttt{PreProcess} class will alter the supplied value of the `page' form variable. If the form fails to validate, for example, or the user has yet to click the submit button, the value of the `page' form variable is left unchanged.
\begin{figure}[H]
\begin{lstlisting}[basicstyle={\sffamily\footnotesize},frame=single,framesep=3pt,xleftmargin=3pt,xrightmargin=3pt]
function PreProcess ( array props ) {
    switch ( props{ "FORM.page" } ) {
        case "feedback" : {

            FeebackForm form = new FeebackForm();

            switch ( form.preprocess( props ) ) {
                case FeebackForm.SUCCESS :
                    props{ "FORM.page" } = "feedback_success"; break;
                case FeebackForm.FAILURE : 
                    props{ "FORM.page" } = "feedback_failure"; break;
                case FeebackForm.INVLALID : break;
                case FeebackForm.PASSIVE : break;
            }
        }
    }
}
\end{lstlisting}
\caption{The \texttt{PreProcess} class constructor}
\label{PreProcess}
\end{figure}
In the \texttt{preprocess()} method of the \texttt{FeedbackForm} class the outermost conditional statement checks for the presence of the `command' form variable. If this form variable is not present but the \texttt{preprocess()} method has been invoked, the form is being shown for the first time. In this case the \texttt{TemplateForm} class is used to process the form's template and the value of `PASSIVE' is returned. If the form has been submitted on the other hand, it is validated. If invalid, the form's template is processed using the \texttt{TemplateFormErrors} class. In both cases the processed template is held in the global properties array, essentially serialising the form. It is worth noting that these serialised templates have no bearing on the state of a Jasper web application from a theoretical point of view and are effectively ignored from this standpoint, the global properties array merely providing the most convenient place in which to store them. 

To continue, the \texttt{process()} method of the \texttt{FeedbackForm} class retrieves the processed template and returns it. The \texttt{validate()} method checks the form fields and sets the values of the `vExclaim' and `vError' temporary variables if the fields are invalid. The error messages have already been retrieved from a dedicated configuration file by this stage and stored in the global properties array. Their names are prepended with the `ERROR.' string, in order to differentiate them from other types of variable stored therein.

Overall program flow in a basic Jasper web application rarely consists of anything more than preprocessing the appropriate form. The \texttt{PreProcess} class ties each form to a particular page to ensure that the \texttt{preprocess()} method of only one form class is ever invoked. And the hidden `command' form variable, unique to each form, ensures that the form is only validated if need be, and the correct template class is used to process the form template.

%% file: implementation.tex
\vspace{-0.75em}

\section{Implementation}

This section details platform specific considerations. These relate both to configuration of a particular Jasper web application as well as the platform on which it runs.

Listing~\ref{main:PHP} shows the `main' server processs. The language is PHP, but the patterns would be identical for any of the languages that Jasper supports. Before preprocessing, the configuration files are parsed. The first configuration file is specific to PHP and contains just two variables:

\begin{figure}[H]
\begin{lstlisting}[basicstyle={\ttfamily\footnotesize},frame=single,framesep=5pt,xleftmargin=5pt,xrightmargin=5pt]
{
    $props = array();
    
    Form::parse( $props );
    Config::parse( "../../config/php.config", $props );
    Config::parse( "../../config/global.config", $props );
    Config::parseBare( "../../config/error.config", "ERROR", $props );
    
    new PreProcess( $props );
    $templatemain = new TemplateMain();

    header( "Content-type: "."text/html" );
    print( $templatemain->process_file( $props{ "CONFIG.rootDir" }.
                                            "template/".
                                                $props{ "FORM.page" }.
                                                    ".html", $props ) );
}
\end{lstlisting}
\caption{The PHP implementation of the `main' server process}
\label{main:PHP}
\end{figure}

\begin{lstlisting}[basicstyle=\ttfamily\footnotesize]
procPath=php/
procExt=.php
\end{lstlisting}
These ensure that the URI resulting from the `[[MAIN]]' token points correctly to the PHP implememtation of the server process. To recall, the template file to consturct the URL to the `main' server process is:
\begin{lstlisting}[basicstyle=\ttfamily\footnotesize]
[[appPath]][[procPath]]main[[procExt]]
\end{lstlisting}
When processed, the following concrete URI results:
\begin{lstlisting}[basicstyle=\ttfamily\footnotesize]
/php/main.php
\end{lstlisting}
Here the value of the `appPath' variable is `/'. This configuration variable is found in the global configuration file, alongside the `rootDir' configuration variable. Between them, they appraise a Jasper web application of the context in which it sits. 

The Perl implementation of the `main' server process is much the same as the PHP implementation, being as it is an interpreted script. The Java implementation is implemented as a servlet, the entry point being the \texttt{service()} method.
The C\# implementation presents some difficulties. The ASP.net platform does not admit mappings to compiled code in a similar fashion to Java servlets, instead ASPX files are invoked directly via their URLs. It is possible, however, to include no active code, or static content for that matter, in the ASPX file, just a preamble that tells the runtime of the class associated with the ASPX file. The `main' server process can then be implemented in a companion `main.aspx.cs' file, a suitable entry point being the \texttt{OnLoad()} method.





On many systems, the distinction between a web application's server context and the web application itself can become blurred by configuration or other such issues. In Jasper's case, however, the distinction always remains clear. A Jasper web application must be appraised of the server context in which it sits by means of the `appath' and `rootDir' configuration variables, but it is not tied to it in any other way. 

%% file: jms.tex
\section{The Jasper Music Store}

This section introduces the Jasper Music Store
and outlines the changes, suprisingly few, that were made to each of the implementations in order to have them run side by side.

The Jasper Music Store comprises three sites: a store front, secure account pages and an administrative interface. The store front and secure account pages are implemented across each of the four platforms that Jasper supports, making a total of eight web applications. The four store front web applications are functionally identical, only the URLs differ, and it is possible to swap between them at any stage by means of a ``platform bar''. URLs are also generated randomly, both in hyperlinks and in forms, so that the platform constantly changes during the user experience. The secure pages behave similarly.


Forcing the platform to change every time a link is followed or form submitted was simply a question of re-writing the classes responsible for processing the `[[MAIN]]' and `[[MAIN\_SECURE]]' tokens. Firstly, the \texttt{process\_token()} method of the \texttt{TemplateBase} class was adjusted to use methods in the \texttt{Platform} class, shown in listing~\ref{JMS:Platform}. The \texttt{random()} method returns one of the three platforms other than the current one. The above listing was converted from PHP, for example, hence PHP's omission in the \texttt{random()} method. The \texttt{main()} method, passed the random platform, then parses the relevant platform specific configuration file in order to construct the appropriate URL. Note the use of the local properties array in this instance.

\begin{figure}[H]
\begin{lstlisting}[basicstyle={\sffamily\footnotesize},frame=single,framesep=3pt,xleftmargin=3pt,xrightmargin=3pt]
function random () {
    switch ( rand( 1, 3 ) ) {
        case 1: return ( "Java" );
        case 2: return ( "ASP" );
        case 3: return ( "Perl" );
    }    
}

function main ( platform ) {
    array props = array();
    
    Config.parse( "../config/" + strtolower( platform ) + ".config", props );
    Config.parse( "../config/global.config", props );
    
    TemplateHTML templatehtml = new TemplateHTML();

    return ( trim( templatehtml.process_file( props{ "CONFIG.rootDir" } +
                                                  "template/_inc/" +
                                                      "main" +
                                                          ".html", props ) ) );
}

\end{lstlisting}
\caption{The \texttt{Platform} class}
\label{JMS:Platform}
\end{figure}

The one remaining issue was the case of the platform bar being utilised immediately after a form submittal. In many respects this situation is similar to the one of the user refreshing the page immediately after submitting form data. The user might expect to submit a form twice in this case. In fact, Jasper's preprocessing architecture procludes this. To see why this is so, consider the case of the user removing an item from their basket. The relevant parts of the \texttt{PreProcess()} constructor of the store front site are given in listing~\ref{JMS:PreProcess}. Here the \texttt{preprocess()} method of the \texttt{BasketRemoveForm} class is called if the value of the `page' form variable is set to `basket\_remove'. Upon successful submittal of the form data, its value is changed to `basket\_contents'. The \texttt{preprocess()} method of the \texttt{BasketContentsForm} class will then be called, quite deliberately after the \texttt{preprocess()} method of the \texttt{BasketRemoveForm} class, and therefore the result of an item being removed from the shopping basket is that the page is changed to show the new contents of the shopping Basket itself.	

\begin{figure}[H]
\begin{lstlisting}[basicstyle={\sffamily\footnotesize},frame=single,framesep=3pt,xleftmargin=3pt,xrightmargin=3pt]
function PreProcess ( array props ) {
    if ( props{ "FORM.page" } == "basket_remove" ) {
        BasketRemoveForm basketremoveform = new BasketRemoveForm();
        switch ( basketremoveform.preprocess( props ) ) {
            case form.SUCCESS() : {
                props{ "FORM.page" } = "basket_contents"; break;
            }
        }
    }
    if ( props{ "FORM.page" } == "basket_contents" ){
        basketremoveform.preprocess( props );
    }
    PlatformBar.preprocess( props );
}
\end{lstlisting}
\caption{The \texttt{parse()} method of the \texttt{Form} class}
\label{JMS:PreProcess}
\end{figure}

Only once conventional preprocessing is completed is the \texttt{preprocess()} method of the \texttt{PlatformBar} class called, mirroring the form variables in hidden form fields in the forms which make up the buttons in the platform bar. If these form variables are submitted a second time by way of one of these buttons, the `command' form variable is still present, which would under normal circumstances instruct the preprocess method of the \texttt{BasketRemoveForm} class to validate the form data again and attempt to remove the item. However, the value of the `page' form variable has now been changed from `basket\_remove' to `basket\_contents', and therefore the \texttt{preprocess()} method of the BasketRemoveForm class will not be called a second time. The additional form data, including the `command' form variable, is therefore essentially discarded, and the use of the platform bar simply refreshes the page showing the contents of the shopping basket.

It turns out that all the actions implemented during preprocessing follow this pattern and therefore the platform bar works with no amendments at all. In essence, the close binding of the `page' and `command' form variables means that there is no chance of form data being submitted afresh, provided that preprocessing results in the value of the `page' form variable being changed to a value not directly associated with the form in question, which is always the case.

%% file: formal.tex
\section{A formal treatment}

This section gives a description of a formal treatment for Jasper. It is by necessity brief and does not contain detailed definitions of either the semantics of the language adopted or the logic employed in the reasoning. However, the description does go so far as to show what kinds of behaviours can be specified and, taking one of these as a worked example, gives a specification and outlines the verification process. 

The following list shows in plain English what kinds of behaviours can be treated for a simple web application that presents the user with a feedback form and sends an email upon its successful validation: 
\begin{enumerate}
\item a feedback form is shown repeatedly until validated successfully\vspace{-0.25em}
\item the feedback form is validated whenever the submit button is pressed\vspace{-0.25em}
\item appropriate validation errors are shown if the from fails to validate\vspace{-0.25em}
\item an attempt is made to send an email upon successful validation of this form
\end{enumerate}
The listing of the \texttt{process\_token()} method of the \texttt{TemplateFormErrors} class given below demonstrates the language adopted in this treatment. Known as abstract pseudo-code, it is a slimmed down imperative language, a distillation of concrete pseudo-code with many features such as references to configuration and the file system removed:
\begin{myabstractlisting}
&\mathsf{process\_token(}t:string\mathsf{)}\{&\\
&\quad\mathsf{if(token\_name(}t,\mathsf{EXCLAIM))}\{&\\
&\quad\quad\mathsf{if(token\_arg(}t\mathsf{)}=\mathsf{retrieve(VAR.vExclaim))}\{&\\
&\quad\quad\quad\mathsf{return(!)}&\\
&\quad\quad\}&\\
&\quad\quad\mathsf{return(\epsilon)}&\\
&\quad\}&\\
&\quad\mathsf{return(parent.process\_token(}t\mathsf{))}&\\
&\}&
\end{myabstractlisting}
Here, the \texttt{token\_name()} and \texttt{token\_args()} methods act on a token string to retrieve its name and arguments respectively, while the \texttt{retrieve()} method acts on the global properties array, which is considered ubiquitous and is therefore not present as one of the arguments. This listing should be directly comparable with the equivalent concrete pseudo-code listing of the same method or, for that matter, listings in any of the concrete languages in which Jasper is implemented.

Being imperative, this language admits a treatment using Separation Logic~\cite{DBLP:conf/lics/Reynolds02}, which describes the operation of part of a program in terms of the effect it has on the program state. Logical formulae relating to this state before and after execution of part of a program, known as the pre-condition and post-condition respectively, sit either side of that part to form what is known as a Hoare triple~\cite{DBLP:journals/cacm/Hoare09}. A partial specification of the the method listed above is given below, for example:
\[
\langle\mathsf{V_n}\rangle\;\mathsf{TemplateFormErrors.process\_token(EXCLAIM:comments)}\;\langle\mathsf{V_n}\wedge(return=\mathsf{!})\rangle
\]
Here, the argument $t$ has been replaced with a specific value, and the following abbreviation has been used:
\[
\mathsf{V_n}=\{(\mathsf{VAR.vExclaim,\mathsf{comments}})\}
\]
The expression on the right represents the global properties array, with curly braces being chosen to mimic the literals for associative arrays found in several commonplace concrete languages. The ordered pair inside the braces represents an assertion about an entry in this array, in this case the `vExclaim' temporary variable. Because of the use of curly braces in this fashion, angle braces are preferred to enclose pre-conditions and post-conditions to avoid any confusion.

The pre-condition and post-condition here are typical of the kinds found in this treatment. Although the logic includes many of the standard characteristics of first order logics dealing with imperative languages, in practice a significant number of specifications require no more than assertions about entries in the global properties array or the return values of methods. Given that the global properties array holds the values of form and temporary variables, which respectively dicate the behaviour of a server process and characterise its output, perhaps this is not suprising.

Returning to the list of functionality given at the start of this section, the specification of the \texttt{main()} method of the server process for the simple web application described encompasses each of the list's items. Specifications along the lines of the one given already can be given for each item, which in turn form part of the overall specification. Consider again one of these plain English specifications, \textit{``appropriate validation errors are shown if the from fails to validate''} and, as a worked example, the following closely related formal specification:
\[
\langle\mathsf{F_{p(f),c(F),n}}\rangle\;\mathsf{main()}\;\langle\mathsf{F_{p(f),c(F),n}}\wedge(return=f'_\mathsf{f})\rangle
\]
This specification states that given the absence of the `comments' form variable, the associated form field is flagged as invalid. Again the specification contains assertions about entries in the global properties array and, in this case, a return value which is in fact the processed HTML content returned:
\[
\mathsf{F_{p(f),c(F),n}}=\{(\mathsf{FORM.page},\mathsf{feedback}),(\mathsf{FORM.command},\mathsf{FEEDBACK}),(\mathsf{FORM.name},\mathsf{\_})\}
\]
\[
\begin{myarray}[0pt]{rl}
f'_{\mathsf{f}}=[&name\!:\mathsf{(name,)}\\
&comments\!:\mathsf{(comments,)}\;!,\\
&\mathsf{(page,feedback)}\;\mathsf{(command,FEEDBACK)}]
\end{myarray}
\]
In the second abbreviation, ordered pairs represent input fields. Note the presence of the exclamation mark next to the `comments' input field. This has replaced an instance of the `[[EXCLAIM:comments]]' token during processing by to the \texttt{process\_token()} method, the listing of which was shown earlier.

It is evident that the key to a formal treatment is the ability to specify outgoing, character based content and to link this to the global properties array. The remainder of this section therefore details the treatment of the double while loop used in the template classes to process template files, which reults in the global properties array affecting output. To begin with, file and line types are defined:
\[
F_i^M=(i,[l_1,...l_M])\quad l_i=\sum_{j=1}^{N-1}(s_{ij}+t_{ij})+s_N\quad l'_i=\sum_{j=1}^{N-1}(s_{ij}+t'_{ij})+s_N
\]
Here $F_i$ is a file with file pointer $i$ and content a sequence of lines $[l_1,...l_M]$, while $l_i$ is a line consisting of a sequence of tokens delimited by strings. The equivalent processded line $l'_i$ is also shown. Note that the tokens $t'_{ij}$ have been processed but the delimiting strings $s_{ij}$ remain the same. To continue, the \texttt{process\_file()} method is used to process the contents of a template file line by line. It is always called against an instance of a template class and so its specification includes an arbitrary class instance $c$:
\[
\langle(f=F_1^M)\rangle
\;c.\mathsf{process\_file(}*\myspace f:file\mathsf{)}\;
\langle(f=F_0^M)\wedge(return=[l'_1,...l'_M])\rangle
\]
Within the body of this method's while loop the \texttt{process\_line()} method is called against the `context' keyword, which is similar to the `this' keyword in many languages, and its specification is given next:
\[
\langle(context=c)\rangle
\;\mathsf{context.process\_line(}l_i:line\mathsf{)}\;
\langle(context=c)\wedge(return=l'_i)\rangle
\]
In a similar fashion, within the body of this method's while loop the \texttt{process\_token()} method of the class an instance of which the \texttt{process\_file()} method was initially called against is called by way of the `context' keyword. Processed tokens are defined below, with class inhertence treated by ordering the classes, with $d$ effectively being the first subclass of $c$ which contains a \texttt{process\_token()} method, or $c$ itself:
\[
t'_{ij}=d.\mathsf{process\_token(}t_{ij}\mathsf{)}\quad d=\max\{d'|d'\leq c\}
\]
Loop conditions can be formulated to allow the while loops to be verified and the result is the specification of the \texttt{process\_file()} method given earlier. The verification of this method tpyically forms the last step in the verification of the main \texttt{main()} method of the server process for the web application.

To sum up, in the worked example the presence of an exclamation mark against an invalid field has been verified. Each such partial specification is in fact just part of the specification of \texttt{main()} method and therefore it is possible to verify the behaviour of the web application as a whole. Such proofs are at the moment ``pen and paper'' but are nonetheless extremely useful.

%% file: related.tex
\section{Related work}

This section outlines a few of the more recent web programming frameworks with theoretical aspects, explaining some of their main features and drawing comparisons with Jasper.

The AFAX project~\cite{Petr(íc(ek_afax:rich} uses F\# to build single-language, type-checked Ajax applications. F\# is executed natively on the server whilst a subset of the language is translated to JavaScript on the client side. Ur~\cite{DBLP:conf/osdi/Chlipala10}, a domain specific, statically typed functional language, allows well-typed programs that won't return malformed HTML, for example, or make extraneous Ajax calls. Further formal treatments in Ur's case include static checking of security policies, expressed as SQL queries.

Hop~\cite{DBLP:conf/mm/Serrano07} is a domain specific, stratified language for writing Ajax applications that is not database-centric, but is instead designed for web applications with graphical user interfaces that require frequent communication between web server and client. It enforces a strict separation between application and presentation logic by way of a dual execution model. The syntax of the Hop language itself can be described as a programmatic variation on HTML. Expressions can be nested inside HTML elements to provide the values for attributes, for example, but also more powerful GUI-led functionality, such as dynamic drop down lists.

Opa~\cite{opa} is both a web application programming platform and a language. Both client and server code are written in the same statically typed language, which bears similarities to C and JavaScript. It is possible to write a complete web application without making the distintiction between the client and server code and, furthermore, the language also supports NoSQL databases. WebDSL~\cite{DBLP:conf/gttse/Visser07} is a domain specific language designed specifically for the web with a rich data model which supports multiple tiers, or aspects, of a web application.

Looking at the available literature, some significant parallels can be found with Jasper. For example, the Mawl project~\cite{DBLP:journals/tse/AtkinsBBC99} introduced the notion of a template. The $<$bigwig$>$ project~\cite{DBLP:journals/toit/BrabrandMS02} introduced the page centred and script centred approaches, the former signifying the placement of code directly in resources, usually HTML files, whilst the latter signifies a template based approach. And, just as in Jasper's case, the page centred approach was discarded in favour of script centred approach for any but the most basic web services. Finally, JWIG shared some similarities with Jasper. It was implemented in a concrete language, for example, and included a framework with which to build web services.

Beyond these similarities, however, there are several important differences. Jasper does not encompass a domain specific language, for example, instead supporting several concrete web programming languages directly. And by supporting several web programming platforms directly, Jasper can claim a measure of platform independence that the other frameworks cannot. 

%% file: conclusions.tex
\section{Conclusions and future work}

In this paper the Jasper web programming framework has been introduced. The conceptual steps that led to its particular design have been elucidated and the ways in which it can be extended in order to build a basic web application have been described. Its essentially platform independent nature has also been demonstrated, perhaps best evinced by the Jasper Music Store. And a brief decription of a formal treatment has been given.

Jasper is primarily about servicing HTTP requests but perhaps what could have be made more evident is the fact Jasper is suited to web services as well as web applications. Although the focus here has been on HTML, for example, other text formats such as JSON and XML can be supported. A better appraisal of other frameworks, both theoretical and commercial, could also have been given. Jasper's core classes are in some sense no more than the distillation of many other simple template frameworks, for example, and this could have been made clearer.

Future work will include a survey of problems related to both web applications and web services. By proposing a set of problem specifications, frameworks that admit a formal treatment, including Jasper, can be compared directly. A formalism for constructing these specifications needs to be found and then suitable specifications given. An example of such a problem is imperviousness to SQL injection attack. The Ur web framework solves this problem as a direct result of functional correctness~\cite{ur}; there is an informal proof that Garble~\cite{garble}, a social gateway powered by Jasper, is secure aganist certain types of SQL injection attack; and recent work on Hop~\cite{DBLP:conf/tosca/LuoRS11} also follows along these lines.

Also planned is a tool called Jape, which will translate from the concrete languages in which Jasper is implemented to concrete pseudo-code and back again. There seems little precedent for this kind of tool in the literature but there is a very successful real world example, namely Facebook's HipHop~\cite{hiphop}, which translates from PHP to C++. And in a recent development, the possibility of replacing concrete pseudo-code with a real-world language has been investigated. The best possible candidate appears to be Dafny~\cite{DBLP:conf/lpar/Leino10}, which is both a language and a program verifier. It does not yet contain all the features that a web language needs, such as class inheritance and support for strings, but these can be added. 
Programmers like tools, not documentation, and it is hoped that these tools will bring the formal treatment of web applications to a much wider audience of web programmers at large.

The author kindly acknowledges advice and support from Sergio Maffeis.

%% file: wwv.bbl
\begin{thebibliography}{10}
\providecommand{\bibitemdeclare}[2]{}
\providecommand{\surnamestart}{}
\providecommand{\surnameend}{}
\providecommand{\urlprefix}{Available at }
\providecommand{\url}[1]{\texttt{#1}}
\providecommand{\href}[2]{\texttt{#2}}
\providecommand{\urlalt}[2]{\href{#1}{#2}}
\providecommand{\doi}[1]{doi:\urlalt{http://dx.doi.org/#1}{#1}}
\providecommand{\bibinfo}[2]{#2}

\bibitemdeclare{misc}{opa}
\bibitem{opa}
\emph{\bibinfo{title}{Opa Web application programming platform}}.
\newblock \bibinfo{howpublished}{\url{http://opalang.org/}}.

\bibitemdeclare{article}{DBLP:journals/tse/AtkinsBBC99}
\bibitem{DBLP:journals/tse/AtkinsBBC99}
\bibinfo{author}{David~L. \surnamestart Atkins\surnameend},
  \bibinfo{author}{Thomas \surnamestart Ball\surnameend},
  \bibinfo{author}{Glenn \surnamestart Bruns\surnameend} \&
  \bibinfo{author}{Kenneth~C. \surnamestart Cox\surnameend}
  (\bibinfo{year}{1999}): \emph{\bibinfo{title}{Mawl: A Domain-Specific
  Language for Form-Based Services}}.
\newblock {\sl \bibinfo{journal}{IEEE Trans. Software Eng.}}
  \bibinfo{volume}{25}(\bibinfo{number}{3}), pp. \bibinfo{pages}{334--346}.
\newblock \urlprefix\url{http://dx.doi.org/10.1109/32.798323}.

\bibitemdeclare{inproceedings}{DBLP:conf/tldi/BaltopoulosG09}
\bibitem{DBLP:conf/tldi/BaltopoulosG09}
\bibinfo{author}{Ioannis~G. \surnamestart Baltopoulos\surnameend} \&
  \bibinfo{author}{Andrew~D. \surnamestart Gordon\surnameend}
  (\bibinfo{year}{2009}): \emph{\bibinfo{title}{Secure Compilation of a
  Multi-Tier Web Language}}.
\newblock In: {\sl \bibinfo{booktitle}{TLDI}}, pp. \bibinfo{pages}{27--38}.
\newblock \urlprefix\url{http://dx.doi.org/10.1145/1481861.1481866}.

\bibitemdeclare{article}{DBLP:journals/toit/BrabrandMS02}
\bibitem{DBLP:journals/toit/BrabrandMS02}
\bibinfo{author}{Claus \surnamestart Brabrand\surnameend},
  \bibinfo{author}{Anders \surnamestart M{\o}ller\surnameend} \&
  \bibinfo{author}{Michael~I. \surnamestart Schwartzbach\surnameend}
  (\bibinfo{year}{2002}): \emph{\bibinfo{title}{The $<$bigwig$>$ project}}.
\newblock {\sl \bibinfo{journal}{ACM Trans. Internet Techn.}}
  \bibinfo{volume}{2}(\bibinfo{number}{2}), pp. \bibinfo{pages}{79--114}.
\newblock \urlprefix\url{http://dx.doi.org/10.1145/514183.514184}.

\bibitemdeclare{misc}{ur}
\bibitem{ur}
\bibinfo{author}{Adam \surnamestart Chlipala\surnameend}:
  \emph{\bibinfo{title}{The Ur Programming Language Family}}.
\newblock \bibinfo{howpublished}{\url{http://www.impredicative.com/ur/}}.

\bibitemdeclare{inproceedings}{DBLP:conf/osdi/Chlipala10}
\bibitem{DBLP:conf/osdi/Chlipala10}
\bibinfo{author}{Adam \surnamestart Chlipala\surnameend}
  (\bibinfo{year}{2010}): \emph{\bibinfo{title}{Static Checking of
  Dynamically-Varying Security Policies in Database-Backed Applications}}.
\newblock In: {\sl \bibinfo{booktitle}{OSDI}}, pp. \bibinfo{pages}{105--118}.

\bibitemdeclare{misc}{hiphop}
\bibitem{hiphop}
\bibinfo{author}{\surnamestart Facebook\surnameend}:
  \emph{\bibinfo{title}{HipHop}}.
\newblock
  \bibinfo{howpublished}{\url{https://developers.facebook.com/blog/post/358/}}.

\bibitemdeclare{inproceedings}{DBLP:conf/sigmod/GernerYDGRS06}
\bibitem{DBLP:conf/sigmod/GernerYDGRS06}
\bibinfo{author}{Nicholas \surnamestart Gerner\surnameend},
  \bibinfo{author}{Fan \surnamestart Yang\surnameend}, \bibinfo{author}{Alan~J.
  \surnamestart Demers\surnameend}, \bibinfo{author}{Johannes \surnamestart
  Gehrke\surnameend}, \bibinfo{author}{Mirek \surnamestart
  Riedewald\surnameend} \& \bibinfo{author}{Jayavel \surnamestart
  Shanmugasundaram\surnameend} (\bibinfo{year}{2006}):
  \emph{\bibinfo{title}{Automatic client-server partitioning of data-driven web
  applications}}.
\newblock In: {\sl \bibinfo{booktitle}{SIGMOD Conference}}, pp.
  \bibinfo{pages}{760--762}.
\newblock \urlprefix\url{http://dx.doi.org/10.1145/1142473.1142580}.

\bibitemdeclare{article}{DBLP:journals/cacm/Hoare09}
\bibitem{DBLP:journals/cacm/Hoare09}
\bibinfo{author}{C.~A.~R. \surnamestart Hoare\surnameend}
  (\bibinfo{year}{2009}): \emph{\bibinfo{title}{Viewpoint - Retrospective: an
  axiomatic basis for computer programming}}.
\newblock {\sl \bibinfo{journal}{Commun. ACM}}
  \bibinfo{volume}{52}(\bibinfo{number}{10}), pp. \bibinfo{pages}{30--32}.

\bibitemdeclare{inproceedings}{DBLP:conf/lpar/Leino10}
\bibitem{DBLP:conf/lpar/Leino10}
\bibinfo{author}{K.~Rustan~M. \surnamestart Leino\surnameend}
  (\bibinfo{year}{2010}): \emph{\bibinfo{title}{Dafny: An Automatic Program
  Verifier for Functional Correctness}}.
\newblock In: {\sl \bibinfo{booktitle}{LPAR (Dakar)}}, pp.
  \bibinfo{pages}{348--370}.
\newblock \urlprefix\url{http://dx.doi.org/10.1007/978-3-642-17511-4_20}.

\bibitemdeclare{inproceedings}{DBLP:conf/tosca/LuoRS11}
\bibitem{DBLP:conf/tosca/LuoRS11}
\bibinfo{author}{Zhengqin \surnamestart Luo\surnameend},
  \bibinfo{author}{Tamara \surnamestart Rezk\surnameend} \&
  \bibinfo{author}{Manuel \surnamestart Serrano\surnameend}
  (\bibinfo{year}{2011}): \emph{\bibinfo{title}{Automated Code Injection
  Prevention for Web Applications}}.
\newblock In: {\sl \bibinfo{booktitle}{TOSCA}}, pp. \bibinfo{pages}{186--204}.

\bibitemdeclare{misc}{Petr(íc(ek_afax:rich}
\bibitem{Petr(íc(ek_afax:rich}
\bibinfo{author}{Tomᚠ\surnamestart Petr(íc(ek\surnameend} \&
  \bibinfo{author}{Don \surnamestart Syme\surnameend}:
  \emph{\bibinfo{title}{AFAX: Rich client/server web applications in F\#}}.

\bibitemdeclare{inproceedings}{DBLP:conf/lics/Reynolds02}
\bibitem{DBLP:conf/lics/Reynolds02}
\bibinfo{author}{John~C. \surnamestart Reynolds\surnameend}
  (\bibinfo{year}{2002}): \emph{\bibinfo{title}{Separation Logic: A Logic for
  Shared Mutable Data Structures}}.
\newblock In: {\sl \bibinfo{booktitle}{LICS}}, pp. \bibinfo{pages}{55--74}.
\newblock
  \urlprefix\url{http://doi.ieeecomputersociety.org/10.1109/LICS.2002.1029817}.

\bibitemdeclare{inproceedings}{DBLP:conf/mm/Serrano07}
\bibitem{DBLP:conf/mm/Serrano07}
\bibinfo{author}{Manuel \surnamestart Serrano\surnameend}
  (\bibinfo{year}{2007}): \emph{\bibinfo{title}{Programming web multimedia
  applications with Hop}}.
\newblock In: {\sl \bibinfo{booktitle}{ACM Multimedia}}, pp.
  \bibinfo{pages}{1001--1004}.
\newblock \urlprefix\url{http://dx.doi.org/10.1145/1291233.1291450}.

\bibitemdeclare{misc}{garble}
\bibitem{garble}
\bibinfo{author}{James \surnamestart Smith\surnameend}:
  \emph{\bibinfo{title}{The Garble personalised social gateway}}.
\newblock \bibinfo{howpublished}{\url{http://www.mygarble.com/}}.

\bibitemdeclare{misc}{jms}
\bibitem{jms}
\bibinfo{author}{James \surnamestart Smith\surnameend}:
  \emph{\bibinfo{title}{Jasper Music Store}}.
\newblock \bibinfo{howpublished}{\url{http://www.jaspermusicstore.com/}}.

\bibitemdeclare{misc}{jasper}
\bibitem{jasper}
\bibinfo{author}{James \surnamestart Smith\surnameend}:
  \emph{\bibinfo{title}{The Jasper website}}.
\newblock \bibinfo{howpublished}{\url{http://jasper.mygarble.com/}}.

\bibitemdeclare{misc}{jasper-api}
\bibitem{jasper-api}
\bibinfo{author}{James \surnamestart Smith\surnameend}:
  \emph{\bibinfo{title}{The Jasper website API Documentation}}.
\newblock \bibinfo{howpublished}{\url{http://jasper.mygarble.com/article/api}}.

\bibitemdeclare{inproceedings}{DBLP:conf/gttse/Visser07}
\bibitem{DBLP:conf/gttse/Visser07}
\bibinfo{author}{Eelco \surnamestart Visser\surnameend} (\bibinfo{year}{2007}):
  \emph{\bibinfo{title}{WebDSL: A Case Study in Domain-Specific Language
  Engineering}}.
\newblock In: {\sl \bibinfo{booktitle}{GTTSE}}, pp. \bibinfo{pages}{291--373}.
\newblock \urlprefix\url{http://dx.doi.org/10.1007/978-3-540-88643-3_7}.

\end{thebibliography}
